\documentclass[%
 aip,
 amsmath,amssymb,
 reprint,%
]{revtex4-1}

\usepackage{natbib}
\usepackage{graphicx}
\usepackage{bm}
\usepackage{color,soul}
\usepackage{hyperref}
\usepackage{float}
\usepackage{csquotes}
\usepackage{color,soul}
\usepackage{hyperref}
\hypersetup{colorlinks=true, urlcolor=blue, citecolor=blue}

\begin{document}

\preprint{APS}
\title{A Generic Slater-Koster Description of the Electronic Structure of Centrosymmetric Halide Perovskites \\
 }


\author{Ravi Kashikar}
 \author{Mayank Gupta}
\author{B. R. K. Nanda}
\email{nandab@iitm.ac.in}

\affiliation{
  Condensed Matter Theory and Computational Lab, Department of Physics,
Indian Institute of Technology Madras, Chennai - 36, India
}

\date{\today}
\begin{abstract}

The halide perovskites have truly emerged as efficient optoelectronic materials and show the promise of exhibiting nontrivial topological phases. Since the bandgap is the deterministic factor for these quantum phases, here we present a comprehensive electronic structure study using first-principle methods by considering nine inorganic halide perovskites CsBX$_3$ (B = Ge, Sn, Pb; X = Cl, Br, I) in their three structural polymorphs (cubic, tetragonal and orthorhombic).  A series of exchange-correlations (XC) functionals are examined towards accurate estimation of the bandgap. Furthermore, while thirteen orbitals are active in constructing the valence and conduction band spectrum, here we establish that a four orbital based minimal basis set is sufficient to build the Slater-Koster tight-binding model (SK-TB), which is capable of reproducing the bulk and surface electronic structure in the vicinity of the Fermi level. Therefore, like the Wannier based TB model, the presented  SK-TB model can also be considered as an efficient tool to examine the bulk and surface electronic structure of halide family of compounds. As estimated by comparing the model study and DFT band structure, the dominant electron coupling strengths are found to be nearly independent of XC functionals, which further establishes the utility of the SK-TB model. 

\end{abstract}
\maketitle

\section{Introduction}

\begin{figure}
\centering
\includegraphics[angle=-0.0,origin=c,height=5.4cm,width=9.0cm]{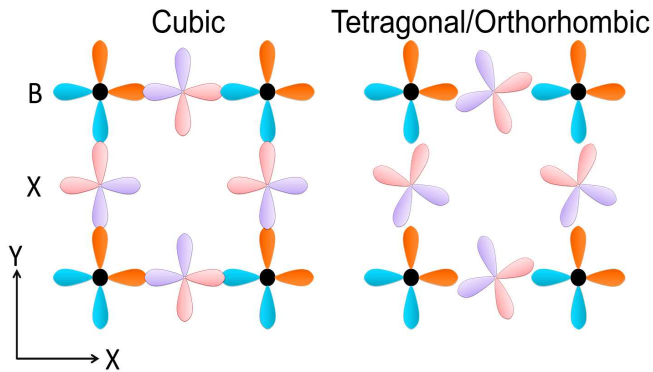}
\caption{Schematic representation of chemical bonding of B-\{s, p\}-X-p orbitals in various polymorphs of halide perovskites.   }
\label{fig1}
\end{figure}

Halide perovskites of the form ABX$_3$ (where A is an organic or inorganic monovalent entity, B is a divalent cation such as Pb, Sn, Ge and X is a halogen (Cl, Br, I and F)), have brought a paradigm shift in the photovoltaic applications because of parity allowed direct transitions between the band extrema \cite{1,2,3,4}. In recent times along with the optical properties, these perovskites have shown the ferroelectrically driven spin-texture, and topological quantum phase transition under external forces in both centrosymmetric and noncentrosymmetric phases \cite{Nature_mat, Nano_letter, 32, Jin, doping2, strain4, 21,22}. While the entity A predominantly decides the structural stability  and centrosymmetricity, B and X governs the electronic properties of these compounds \cite{31,32}. The cubic phase of halide perovskites exhibit structural phase transition with temperature and pressure, and the resulted lower symmetry crystal structures are characterized by in-plane as well as out-of-plane octahedral rotations \cite{41,struct1,struct2}. The schematic representation of the orbital overlap in high symmetric and lower symmetric phases is illustrated in the Fig. \ref{fig1}.  

Over the last decade, both experimental and  density functional theory (DFT) methods have been employed to unravel the orbital and crystal interplay to study the optoelectronic and other intriguing properties of these perovskite materials. The previous electronic structure studies on these compounds suggest that the band spectrum consists of anti-bonding and bonding states, along with the X-p dominated non-bonding states arise out of strong covalent hybridization between B-\{s, p\} and X-p states\cite{Lamb_2, 31,32}. Nature of the band structure remains similar for a particular phase with varying bandwidth for a family of halide perovskites.  Beyond this basic observations, there are several issues which need to be addressed to construct a comprehensive picture of the electronic structure of this important class of compounds. For example choice of exchange correlation functional is one of the debatable issue \cite{jishi1,Jishi2}. 

As the bandgap varies widely in this class of compounds and the materials applicability in optoelectronic devices as well as in inducing non-trivial topological phases primarily depend on the bandgap, and we know that the bandgap highly sensitive to type of exchange-correlation functionals employed within the DFT formalism. Therefore, it is pertinent that a relation between the bandgap and the type of exchange correlation functional be established, which could capture the experimental observations. Furthermore, this class of compounds exhibit three temperature dependent structural phases governed by anionic displacements \cite{struct1,struct2}. The bandgap of these phases differ significantly from each other. Therefore, it is crucial to identify the chemical interactions that govern the bandgap in this family. Also, it has been observed that electronic structure is sensitive to both B and X. For example if B is Sn for any X, we observe a lower bandgap as compared Pb and Ge  and the reason has not been able to explain through parameter-free density functional calculations. As a whole comprehensive first principles calculations and formulation of model Hamiltonians are required to provide a generic description of the electronic structure of the halide perovskites.


In the literature, from the model Hamiltonian perspective, only a handful of studies have examined the band structure of halide perovskites. Boyer-Richard et al., envisaged the fourteen orbital basis based tight-binding (TB) Hamiltonian without incorporating the second neighbour interaction and Jin et al. have studied the continuum model based on the four orbital basis TB Hamiltonian \cite{Boyer_Richard, Jin}. In our recent work, the four orbital basis model is employed to lower symmetry polymorphs to examine the band topology of these perovskites systems \cite{32}. However, all these studies were carried out on individual members, and so far, there has been no systematic study of TB Hamiltonian for all the members of the halide perovskites family to understand sheer variety of properties with varying B and X elements. This is the first attempt to bring the generic picture of halide perovskites, and for that, we have considered the nine members and three structural phases to analyze the role of each entity of inorganic ABX$_3$ compounds.


\section{Computational details}

In the present work, we have employed both pseudopotential method with the plane-wave basis set as implemented Vienna Abinitio Simulation Package (VASP), and full-potential linearized augmented plane wave (FP-LAPW) method as implemented in WIEN2k simulation tool for DFT calculations \cite{VASP, LAPW, Blaha}. In both the methods, the Perdew-Burke-Ernzerhof (PBE) exchange-correlation functional within the generalized gradient approximation (GGA) is considered. However, the PBE underestimates the bandgap severely as compared to the experimental bandgap. Therefore, the PBE-GGA approximation along with modified Becke-Johnson (mBJ) potential as well as hybrid functionals are used to take into account the exchange-correlation effect and for comparison purpose \cite{GGA, mbj-1}. All the band structures in the present work are presented from PBE+Tran-Blaha mBJ potential. The self-consistent field (SCF) calculations comprise of augmented plane waves of the interstitial region and localized orbitals: B-\{ns, np\} (B = Ge, Sn, Pb) and  X-p (X = Cl, Br, I). Here, n and m vary with B and X. The R$_{MT}^{X}$ set to 7.0 for all the compounds. The Brillouin zone integration is carried out with a Monkhorst-Pack grid. We used a $k$-mesh of 10$\times$10$\times$10 (yielding 35 irreducible points), 6$\times$6$\times$4 (yielding 35 irreducible points), and 8$\times$8$\times$6 (yielding 100 irreducible points) for the $\alpha$, $\beta$ and $\gamma$-phases respectively. 
To build a Slater-Koster tight binding (SK-TB) Hamiltonian, we consider the appropriate basis set out of the orbitals dominating the bands at the Fermi level and this information is obtained from the first-principle based density functional calculations. The details of the Hamiltonian and orbital basis is discussed in details in section IIIB.

\begin{table*}
    \centering
    \caption{ Structural parameters of various CsBX$_3$ perovskites as obtained from DFT calculations and in comparison with the experimental results. The compounds exist in different crystal polymorphs at different temperature ranges. The $\theta_{ab}$ and $\theta_c$ is 180$^{\circ}$ for $\alpha$ phase.}
    \begin{tabular}{cccccccc}
    \hline
    \hline
   Phase&B&X&\multicolumn{2}{c}{Lattice Parameter (\AA)} &Octahedral angle & Temperature (K)&Ref.\\
    &&&DFT&Expt.& $\theta_{ab}$, $\theta_c$ &\\
    \hline
&&Cl&5.34&5.47&& 443,449 & \cite{gex1, gecl1}  \\
&Ge&Br&5.6&5.69&&543 & \cite{gex1}  \\
&&I&6.0&6.05&&573 & \cite{gex1} \\
&&Cl&5.62&5.55&&293 & \cite{cssncl_br} \\
$\alpha$&Sn&Br&5.88&5.8&& 292,300& \cite{cssnbr} \\
&&I&6.27&6.21&& 426 & \cite{cssni} \\
&&Cl&5.71&5.6& & 320 & \cite{cspbcl} \\
&Pb&Br&5.98&5.87&&403 & \cite{cspbbr}  \\
&&I&6.38&6.29& &554 & \cite{cspbi} \\
\hline

&Sn&Br&8.27	8.27	5.92 &8.18, 8.18, 5.82&  162.8 $^{\circ}$, 180 $^{\circ}$ &270-300 & \cite{cssnbr}\\
$\beta$&&I&8.81	8.81	6.31&8.77, 8.77, 6.26& 162.6 $^{\circ}$, 180 &351-426& \cite{cssni}\\
&Pb&Br&8.35, 8.35, 6.04&& 156.3$^{\circ}$,  180$^{\circ}$ &361-403& \cite{cspbbr} \\
&&I&8.89	8.89	6.44 &8.82, 8.82, 6.3&  155.9 $^{\circ}$, 180$^{\circ}$ &457-554&\cite{cspbi} \\
\hline
&Sn&Br&8.15	8.37	11.77 &8.19, 11.58, 8.02& 157.4 $^{\circ}$, 163.5$^{\circ}$ &270& \cite{cssnbr}  \\
$\gamma$&&I&8.66	8.92	12.53&8.68, 8.64, 12.37&  154.6 $^{\circ}$, 161.4$^{\circ}$ &426& \cite{cssni}\\
&Pb&Br&8.24	8.47	11.93&8.21, 8.25, 11.76& 153.4 $^{\circ}$, 160.6$^{\circ}$ &361& \cite{cspbbr} \\
&&I&8.76	9.04	12.7&8.62, 8.85, 12.5&152.3 $^{\circ}$, 159.2$^{\circ}$ &457& \cite{cspbi} \\
\hline
\hline
    \end{tabular}
    \label{T1}
\end{table*}

\section{Results and Discussion}
\subsection{DFT study}
\begin{figure}
\centering
\includegraphics[angle=0.0,origin=c,height=10cm,width=8.8cm]{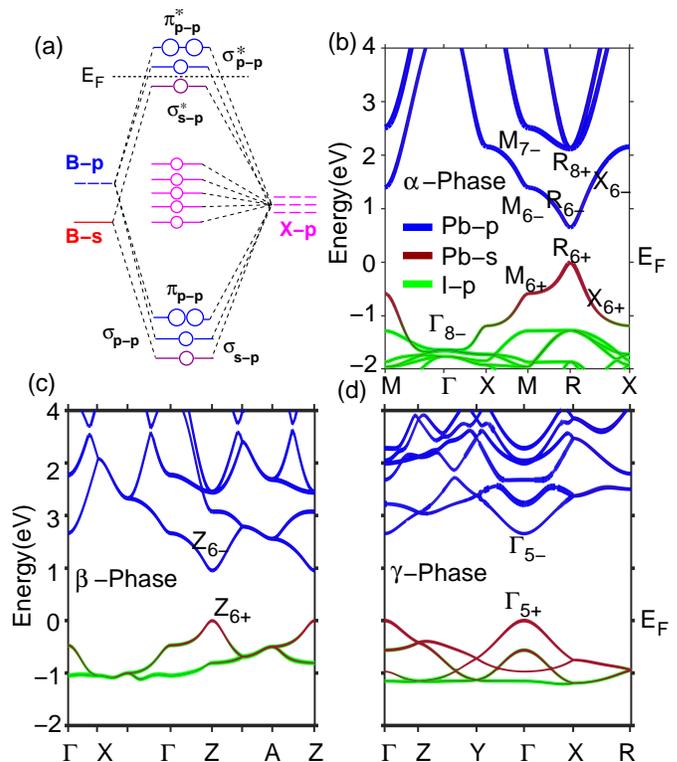}
\caption{ (a) Molecular orbital picture of halide perovskites envisaged from the B-\{s, p\}-X-p atomic orbitals, produces the bonding and antibonding orbitals along with the nonbonding orbitals. (b-d) Band structure of cubic, tetragonal and orthorhombic band structure of representative compound CsPbI$_3$.   }
\label{fig2}
\end{figure}

To start with, we will first discuss the electronic structure of CsPbI$_3$ and explain it through a generic molecular orbital picture as shown in Fig. \ref{fig2}. Here, we outline some of the standard features of band spectrum of halide perovskites: (I) The band spectrum in general consist of four antibonding and bonding bands along with the five nonbonding bands arising due to the  B-\{s, p\}-X-p covalent hybridization in BX$_6$ octahedron.  Thus, the eigenfunctions of antibonding and bonding states are the linear combination of  B-s, B-p and X-p orbitals. Their strength varies with the type of atom at B, halogens as well as crystal symmetry. The nonbonding states are from X-p orbitals. The molecular orbital picture presented in Fig. \ref{fig2}a is for O$_h$ point group symmetry which exists at R and $\Gamma$ points of cubic Brillouin zones. The X-p orbitals undergo symmetry adapted linear combinations and form the molecular orbitals with B-\{s, p\} orbitals. (II) The Fermi level in the electronic spectrum is solely determined by the valence electron count (VEC) which is the total number of valence electrons available per formula unit.   In single halide perovskites, the atom at $A$ contributes one electron, while atoms $B$ and $X$ contributes four and five electrons respectively. Thus, the VEC for halide perovskites turns out to be 20. This makes the states up to $\sigma^*_{s-p}$ occupied and hence the Fermi level lies in the gapped region. (III) All of the inorganic halide perovskites have direct bandgap in nature. The bandgap value varies with the crystal symmetry and chemical composition of ABX$_3$ and spin-orbit coupling (SOC) strength of $B$ site atom.(IV) The parities of the valence band and conduction band edges for three polymorphs in terms of Koster notations is denoted in Fig. \ref{fig2}. The analysis indicates that all the halide perovskites exhibit parity allowed transitions due to the opposite of parity of band edges.

The instability of the high-temperature cubic phase introduces the octahedral rotations in the unitcell, and there occurs the crystal phase transition from the cubic to tetragonal to orthorhombic phases with temperature as listed in Table \ref{T1}. The tetragonal phase is characterized by only inplane octahedral rotation, whereas the orthorhombic phase is characterized both inplane and out of plane octahedral rotations. These rotations increases the unitcell volume as well as the number of atoms.  The Ge based halide perovskites exhibit single phase transition and crystallize in rhombohedral unitcell in room temperature phase.

\begin{figure}
\centering
\includegraphics[angle=-0.0,origin=c,height=3.5cm,width=9cm]{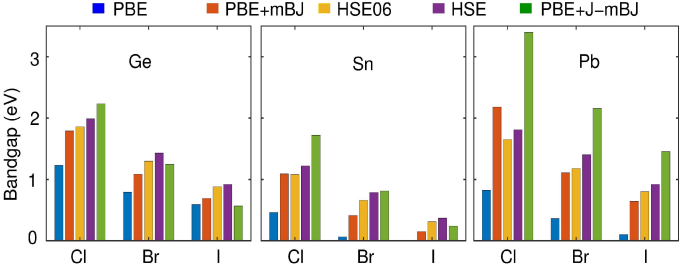}
\caption{Bar chart representation of bandgap of cubic halide perovskites under various exchange correlation functionals. Here, HSE06 and HSE are calculated with pseudopotential methods with mixing parameter 0.25 and 0.3, respectively. }
\label{fig3}
\end{figure}

\begin{figure}
\centering
\includegraphics[angle=-0.0,origin=c,height=7cm,width=7cm]{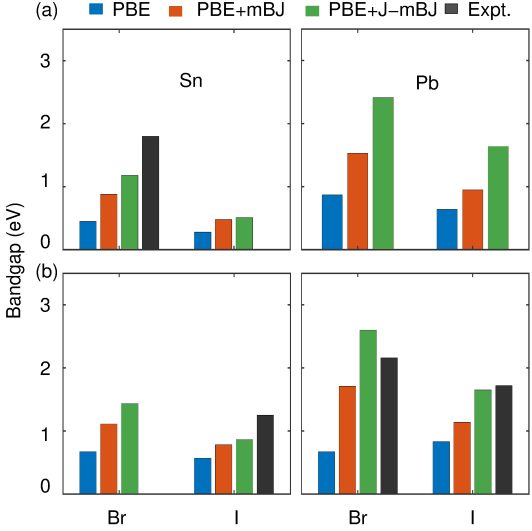}
\caption{Bar chart representation of bandgap of tetragonal (a) and orthorhombic (b) halide perovskites under various exchange correlation functionals. The experimental bandgap values of tetragonal and orthorhombic systems are obtained from \cite{BG_tetra_cssnbr, Nature_comm, cspbi_bg}. }
\label{fig99}
\end{figure}

In Fig. \ref{fig3} and Fig. \ref{fig99} we have compared the bandgap of cubic and lower symmetry halide perovskites for various exchange-correlation functionals. It is known that GGA-PBE underestimates the bandgap significantly, the correction is added through several other approximations. The modified Becke-Johnson (mBJ) proposed by Tran-Blaha improves the bandgap significantly in most of the halide perovskites. The hybrid functionals with mixing parameter $\alpha = 0.25$ (HSE06) and $\alpha = 0.3$ (HSE) offer better estimation of bandgap in few compounds and in others the values are similar to those obtained using TB-mBJ functional. The recently proposed mBJ potential by Jishi et al. provides the bandgap close to experimental one in the case of Pb based compounds \cite{jishi1, Jishi2}. In few cases; CsGeI$_3$ and CsSnI$_3$ hybrid functional minimizes the error as compared to Jishi-mBJ. The bandgap for lower symmetry polymorphs for PBE. mBJ and Jishi-mBJ are shown in Fig. \ref{fig99}, in comparison with available experimental values. It is observed that the gap decreases from Cl to Br to I, irrespective of atoms at A and B sites, and Sn-based halide perovskites have lower bandgap values as compared to the Pb and Ge based perovskites. A broad explanation to this end can be given by comparing the free atomic energy eigenvalues. The difference between  the onsite energies of Sn valence orbitals ($E_{p1/2} - E_{s1/2}$ = 6.35 eV) is lowest as compared to that of Pb (7.04 eV) and Ge (7.53 eV).  Whereas the onsite energy halogen valence $p$ orbital decreases from I (-7.84 eV) to Br (-8.96 eV) to Cl (-9.96 eV) \cite{SS} The bandgap also increases as we move from cubic to tetragonal to orthorhombic phases due to octahedral rotations which decrease the orbitals overlap as shown in Fig. \ref{fig1}. The valence band maximum (VBM) and conduction band minimum (CBM) are predominantly of  B-s and B-p character. The VBM and CBM occurs at R (0.5, 0.5, 0.5), Z (0, 0, 0.5) and $\Gamma$ (0, 0, 0) $k$-points of cubic, tetragonal and orthorhombic Brillouin zone respectively.  
The DFT calculations offers a limited quantitative understanding to establish a generic description of the electronic structure. Specifically, the type and strength of the interactions that govern the valence and conduction bands in the vicinity of the Fermi level needs to be determined. Therefore, in the next section, we build the Slater-Koster based tight-binding Hamiltonian  using a thirteen orbitals (one B-s, three B-p and nine X-p) basis set. Subsequently, the basis set will be reduced to four.

\subsection{Model Hamiltonian for Halide Perovskites}
The appropriate SK-TB Hamiltonian for the family of CsBX$_3$, in the second quantization notation is expressed as  
\begin{equation}\label{1}
H = \sum_{i,m}\epsilon_{im}c_{im}^\dag c_{im} + \sum_{\langle\langle ij \rangle\rangle;m,n}t_{im jn}(c_{im}^\dag c_{jn} + h.c) + \lambda\textbf{L}\cdot\textbf{S}.
\end{equation}
Here, {\it i }({\it j}) and  $\alpha$ ($\beta$ )  are site and the orbitals indices respectively.  The parameters $\epsilon_{i\alpha}$ and $t_{i\alpha j\beta}$ respectively, represent the on-site energy and hopping integrals. The spin-orbit coupling (SOC) is included in the third term of the Hamiltonian with $\lambda$ denoting the SOC strength. The inclusion of SOC doubles the Hilbert space. Adopting a two centre integral approach J. C. Slater and G. F. Koster, expressed the $t_{i\alpha j\beta}$ with direction cosines (DCS) ($l$, $m$, $n$) of the vector joining site ($i$) to other site ($j$) \cite{slater}. For $s$ and $p$ orbitals, which forms the basis for halide perovskites, the generic expressions are provided below 
\begin{align}
    E_{s,s}&=t_{ss\sigma}\\ \nonumber
    E_{s,p_x}&=lt_{sp\sigma}\\ \nonumber
    E_{s,p_y}&=mt_{sp\sigma}\\ \nonumber
    E_{s,p_z}&=nt_{sp\sigma}\\ \nonumber
    E_{p_x,p_x}&=l^2t_{pp\sigma}+(1-l^2)t_{pp\pi}\\ \nonumber
     E_{p_x,p_m}&=lmt_{pp\sigma}-lmt_{pp\pi}
\end{align}
Using Eq. 1 and 2, we now develop  TB Hamiltonian for the halide perovskites. Thus, the spin independent TB Hamiltonian matrix, with the basis set in the order $\{|s^{B}\rangle$, $|p^{B}_{x}\rangle$, $|p^{B}_{y}\rangle$, $|p^{B}_{z}\rangle$,  $|p^{X1}_{x}\rangle$, $|p^{X1}_{y}\rangle$, $|p^{X1}_{z}\rangle$, $|p^{X2}_{x}\rangle$, $|p^{X2}_{y}\rangle$, $|p^{X2}_{z}\rangle$, $|p^{X3}_{x}\rangle$, $|p^{X3}_{y}\rangle$, $|p^{X3}_{z}\rangle$$\}$, 
can be written as

\begin{equation}
H^{FB}_{TB} = \left( \begin{array}{cc}
M_{4\times4}^{B-B}&M_{4\times9}^{B-X}\\\\
(M_{4\times9}^{B-X} )^{\dag}&M_{9\times9}^{X-X}
\end{array}  \right).
\end{equation}
Here, FB indicates the full basis. The individual blocks of this matrix are as follows,
The $M_{4\times4}^{B-B}$ is given by
\begin{equation}
\left( \begin{array}{cccc}
\epsilon_s+h_1(\Vec{k})	& 2it_{sp\sigma}^{B-B}S_x& 2it_{sp\sigma}^{B-B}S_y &2it_{sp\sigma}^{B-B}S_z  \\
-2it_{sp\sigma}^{B-B}S_x	&\epsilon_{p1}+h_2(\Vec{k}) &	0	&0	\\
-2it_{sp\sigma}^{B-B}S_y &0&\epsilon_{p1}+h_3(\Vec{k})&0\\
-2it_{sp\sigma}^{B-B}S_z	&0	&0	&\epsilon_{p1}+h_4(\Vec{k}) \\
\end{array}  \right)
\end{equation}

\begin{equation}
M_{4\times9}^{B-X} = \left( \begin{array}{ccc}
M^{4\times3} M^{4\times3} M^{4\times3}
\end{array}  \right)
\end{equation}

\begin{equation}
M_{4\times3}^{B-X} = \left( \begin{array}{ccc}
t_{sp\sigma}^{B-X}S_x 	&0	&0	\\
t_{pp\sigma}^{B-X}C_x	&0	&0	\\
0&t_{pp\pi}^{B-X}C_x&0\\
0	&0	&t_{pp\pi}^{B-X}C_x	\\
\end{array}  \right)
\end{equation}
From the Fig. \ref{fig2} it is observed that X-p dominated bands are very narrow $(< 1.0$ eV), suggesting negligible X-\{p\}-X-\{p\} second interactions. Hence the block $M_{9\times9}^{X-X}$ can be approximated as
\begin{equation}
M_{9\times9}^{X-X} = \epsilon_{p2} I_{9 \times 9}
\end{equation}
Here, $\epsilon_s$, $\epsilon_{p1}$ and $\epsilon_{p2}$ are on-site energies of B-s, B-p and X-p orbitals respectively. The terms $C_x$ and $S_x$ are short notations for $2\cos(k_xa/2)$ and $2i\sin(k_xa/2)$ respectively. The dispersion term  $g_i$ $(i=1,2,3,4)$, arising from B-B second neighbour interactions are given by
\begin{eqnarray}
h_1(\Vec{k}) &=& 2t_{ss}^{B-B}(cos(k_xa)+cos(k_ya)+cos(k_za)) \nonumber \\
h_2(\Vec{k}) &=& 2t_{pp\sigma}^{B-B}cos(k_xa)+2t_{pp\pi}^{B-B}[cos(k_ya)+cos(k_za)] \nonumber \\
h_3(\Vec{k}) &=& 2t_{pp\sigma}^{B-B}cos(k_ya)+2t_{pp\pi}^{B-B}[cos(k_xa)+cos(k_za)]\\
h_4(\Vec{k}) &=& 2t_{pp\sigma}^{B-B}cos(k_za)+2t_{pp\pi}^{B-B}[cos(k_xa)+cos(k_ya)] \nonumber
\end{eqnarray}

\begin{figure}
\centering
\includegraphics[angle=0.0,origin=c,height=9.5cm,width=8.5cm]{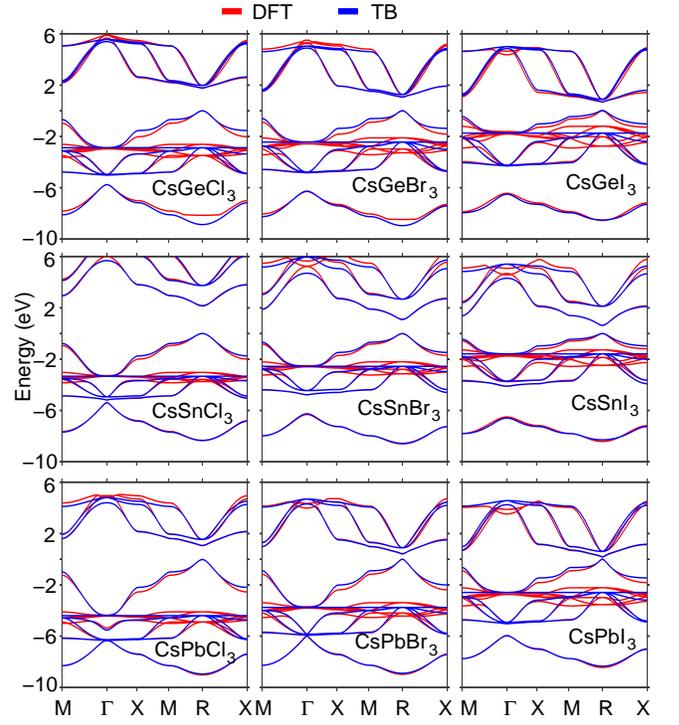}
\caption{ Thirteen orbital basis based TB band structure of CsBX$_3$ in comparison with DFT band structure obtained from GGA+mBJ exchange-correlation functional.  }
\label{fig4}
\end{figure}

The analytical expression for eigenvalues  at time reversal invariant momentum (TRIM) $R$ $(\frac{\pi}{a}, \frac{\pi}{a}, \frac{\pi}{a})$, which is of particular interest as the valence band maximum (VBM) and conduction band minimum (CBM) are observed at this point, and are given as
\begin{align}
E_1[1]& = \frac{(E_p^X + E_s^{B})}{2} -3t_{ss}^{B-B} - \eta \nonumber\\
E_2[8]& = E_p^{X} \nonumber\\
E_3[3]& = E_{p}^{B} - 2t_{pp\sigma}^{B-B} - 4t_{pp\pi}^{B-B} \nonumber\\
E_4[1]& = \frac{(E_p^X + E_s^{B})}{2} -3t_{ss}^{B-B} + \eta.    
\end{align}
Where
\[\eta = \frac{\sqrt{(E_p^{X} - E_s^{B} + 6t_{ss}^{B-B})^2 + 48(t_{sp}^{B-X})^2}}{2}\]
The number in square bracket indicates the degeneracy of the eigenvalues. In presence of SOC, operating $H_{soc} = \lambda \textbf{L.S}$ on B-$p$ orbital basis with the order ($p_{x\uparrow}^{B}$, $p_{y\uparrow}^{B}$, $p_{z\downarrow}^{B}$,$p_{x\downarrow}^{B}$, $p_{y\downarrow}^{B}$, $p_{z\uparrow}^{B}$), we get the following matrix
\begin{equation}
H_{SOC}
 =  \lambda\left( \begin{array}{cccccc}
0&-i&1&&& \\
i&0&-i&&0&\\
1&-i&0&&&\\
&&&0&i&1\\
&0&&-i&0&i\\
&&&1&i&0
\end{array}  \right).
\end{equation} 
  As a case study, the thirteen band model is applied to nine cubic CsBX$_3$ (B = Ge, Sn, Pb; X = Cl, Br, I) perovskites and corresponding bands are fitted with DFT bands to obtain onsite and hopping interactions, which are listed in Table \ref{T3}. The DFT and TB bands are shown in Fig. \ref{fig4}, suggesting the excellent agreement between each other. The table infers that (I) tin-based halide perovskites exhibit higher onsite energies (lowest E$_s$-E$_p$ energy), agrees with the atomic energies mentioned in section-II. A similar trend is observed for halogen orbital energies.  The hopping parameter $t_{sp}^{\sigma}$ varies from Cl to Br to I indicating a decrease in the bandwidth of uppermost valence band for all the compounds as we move from Cl to Br to I. The onsite energies provided in Table \ref{T3} agrees qualitatively well with the recent work of Hoffmann et al. \cite{Hoffmann}. The other hopping parameters $t_{sp}^{B-X}$ and $t_{pp\sigma}^{B-B}$ governs the bandwidth of uppermost valence band and conduction band, and these parameters increase with Cl-Br-I. The other parameters remain more or less constant for all the perovskites.   
\begin{table*}
\centering
\caption{ On-site energy, hopping parameters and GGA-mBJ bandgap of CsBX$_3$ perovskites in units of eV}
\begin{tabular}{cccccccccccccc}
\hline
\hline
B&X&$E_{B-s}$ &$E_{B-p}$ &$E_{X-p}$&$t_{sp}^{B-X}$ &$t_{pp\sigma}^{B-X}$ & $t_{pp\pi}^{B-X}$& $t_{ss}^{B-B}$ & $t_{sp\sigma}^{B-B}$&$t_{pp\sigma}^{B-B}$&$t_{pp\pi}^{B-B}$ & $\lambda$ \\
 \hline

&Cl &-3.17 &	5.35 &	-0.18&	-1.20&	1.94&	-0.58&	0.02&	-0.16&	0.31&	0.03&	0.07\\
Ge&Br&-3.55&	4.76&	0.40&	-1.16&	1.94&	-0.56&	0.02&	-0.15&	0.32&	0.02&	0.07\\
&I&-3.97&	4.29&	0.92&	1.00&	1.92&	-0.52&	0.02&	-0.15&	0.37&	0.02&	0.07\\
&Cl&-0.71&	6.42&	-0.06&	-1.29&	1.94&	-0.52&	-0.08&	-0.2&	0.24&	0.06&	0.16\\
Sn&Br&-1.43&	5.71&	0.36&	-1.27&	1.96&	-0.53&	-0.07&	-0.19&	0.34&	0.05&	0.16\\
&I&-2.34&	4.79&	0.92&	-1.12&	1.9&	-0.53&	-0.02&	-0.17&	0.38&	0.01&	0.14\\
&Cl&-1.61&	7.63&	0.25&	-1.18&	1.88&	-0.57&	-0.03&	-0.13&	0.31&	0.06&	0.53\\
Pb&Br&-3.15&	5.84&	0.43&	-1.13&	1.86&	-0.53&	-0.02&	-0.12&	0.27&	0.04&	0.53\\
&I&-4.11&	4.75&	0.96&	-0.94&	1.82&	-0.45&	0.01&	0.12&	0.25&	0.02&	0.5\\
\hline
 \hline
\end{tabular}
\label{T3}
\end{table*}

\begin{figure}
\centering
\includegraphics[angle=-0.0,origin=c,height=6.0cm,width=6.0cm]{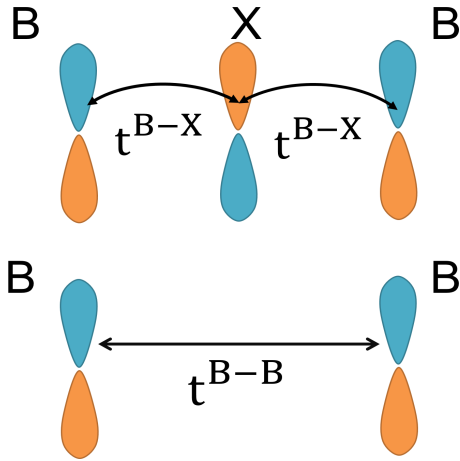}
\caption{Schematic representation hopping interaction in thirteen orbital basis based TB model and four orbital basis based TB model. Here, Fermi level is set to zero.  }
\label{fig5}
\end{figure}

\begin{table}
    \centering
    \caption{ Orbital weight in-terms of \% at conduction band minimum and valence band maximum of different halide perovskites. }
    \begin{tabular}{ccccc}
    \hline
    \hline
  &&CBM&\hspace{0.3cm} VBM\\
    B&X&B-p&B-s&X-p\\
    \hline
&Cl&100&65&35  \\
Ge&Br&100&63&37  \\
&I&100&65&35  \\
&Cl&100&69&31  \\
Sn&Br&100&67&33  \\
&I&100&67&33  \\
&Cl&100&66&34  \\
Pb&Br&100&61&39  \\
&I&100&60&40  \\
\hline
\hline
    \end{tabular}
    \label{T4}
\end{table}

Having understood about the full band spectrum of cubic halide perovskites, now we aim to apply for lower symmetry polymorphs.  However, looking at the size and number atoms in the lower symmetry crystal structures, it is difficult to trace and analyze a large number of interactions in these systems. Therefore, it is necessary to develop minimal basis Hamiltonian having less number of interaction without losing essential physics. The bands forming the VBM and CBM at high symmetry points are primarily created by the four B-\{s, p\} orbitals. From our calculations, we found that the contribution B-p orbital at conduction band minimum is 100\%. Whereas the valence band maximum is made up of a linear combination of X-p and B-s characters as listed in Table \ref{T4}. As we can see, the contribution of X-p orbitals is approximately half of B-s. Thus, there is room to minimize the number of interactions. Therefore, we choose the basis set from B-\{s, p\}, where the interactions between B-X-B will be included in B-B interactions as shown in Fig. \ref{fig5}.  Thus, the SOC incorporated four-band TB Hamiltonian in the matrix form can be written as

\begin{equation}
H_{TB}^{MB}
 =  \left( \begin{array}{cc}
    H_{\uparrow\uparrow}& H_{\uparrow\downarrow} \\
    H_{\downarrow\uparrow}^{\dagger}&  H_{\downarrow\downarrow}
\end{array}  \right), 
\end{equation}

\begin{equation}
\footnotesize
 H_{{\uparrow\uparrow}} 
 =  \left( \begin{array}{cccc}
    \epsilon_s+h_1(\Vec{k}) & 2i(t_{sp}^x)^{AA}S_x & 2i(t_{sp}^x)^{AA}S_y & 2i(t_{sp}^z)^{AA}S_z\\
    -2i(t_{sp}^x)^{AA}S_x & \epsilon_{p}^x+h_2(\Vec{k}) & -i\lambda &0\\ 
    -2i(t_{sp}^x)^{AA}S_y & i\lambda & \epsilon_{p}^x+h_3(\Vec{k}) & 0\\
    -2i(t_{sp}^z)^{AA}S_z &   0 & 0 & \epsilon_{p}^z+h_4(\Vec{k}) 
\end{array}  \right)
\end{equation}

\begin{figure}
\centering
\includegraphics[angle=0.0,origin=c,height=9.5cm,width=8.9cm]{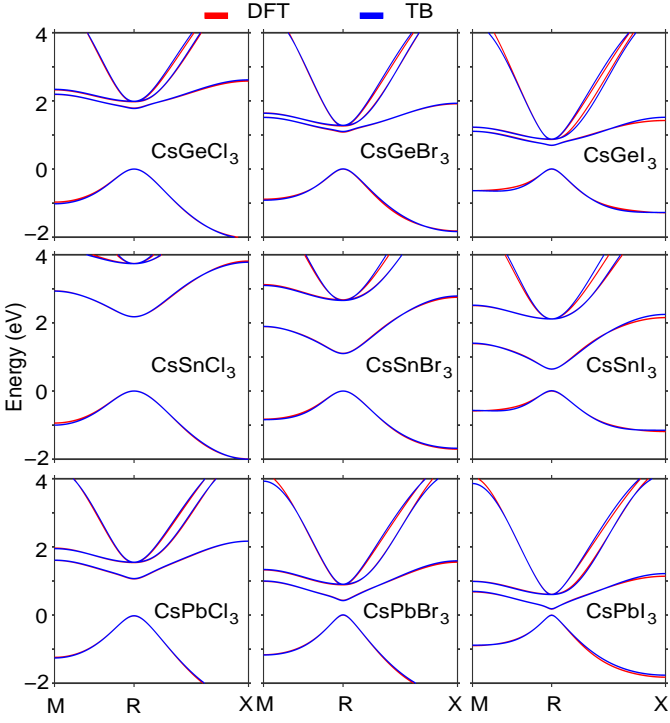}
\caption{ Band structure of various halide perovskites obtained from four orbital based TB Hamiltonian and compared with the DFT band structure. Here, Fermi level is set to zero.  }
\label{fig6}
\end{figure}

Here, MB refers to minimal basis. The $\epsilon^A$s are band centres of the anti-bonding bands, and $t^A$s are the second neighbour hopping integrals. The dispersion functions $f_i$ ($i = 1, 2, 3, 4)$ are expressed in Eq. 8. The TB bands obtained from this four-band model are shown in the Fig. \ref{fig6} for nine cubic halide perovskites and fitting parameters are listed in Table \ref{T5}, and they completely agree with DFT bands along the broad M-R-X $k$-path of the cubic Brillouin zone. Thus, it validates the minimal basis set based model which captures the essential features of halide perovskites around the Fermi level with the less parametric quantities. In the next section, we further validate this minimal basis set based TB model to lower symmetry polymorphs.

\begin{table}[h]
\centering
\caption{ Interaction parameters ($\epsilon^A$s and $t^A$s) and SOC strength $\lambda$ in units of eV.}
\begin{tabular}{cccccccccc}
\hline
\hline
B&X & $\epsilon_s$ & $\epsilon_p$ & $t_{ss}$ & $t_{sp}$& $t_{pp\sigma}$ & $t_{pp\pi}$ & $\lambda$  \\
 \hline
&Cl&1.17 &	6.47&	-0.26&	0.47&	0.75&	0.09&	0.07\\
Ge&Br&1.46&	6.10&	-0.23&	0.48&	0.84&	0.09&	0.06\\
&I&1.70&	5.55&	-0.16&	0.48&	0.86&	0.09&	0.06\\
&Cl&2.08&	8.68&	-0.25&	0.45&	0.74&	0.10&	0.52\\
Sn&Br&1.73&	7.11&	-0.21&	0.50&	0.77&	0.11&	0.52\\
&I&1.68&	6.23&	-0.15&	0.48&	0.83&	0.10&	0.49\\
&Cl&2.46&	7.57&	-0.31&	0.49&	0.72&	0.10&	0.16\\
Pb&Br&2.36&	6.88&	-0.30&	0.52&	0.79&	0.11&	0.16\\
&I&2.18&	6.05&	-0.22&	0.48&	0.85&	0.10&	0.14\\
\hline
\hline
\end{tabular}
\label{T5}
\end{table}

\begin{figure*}
\centering
\includegraphics[angle=0.0,origin=c,height=9cm,width=15cm]{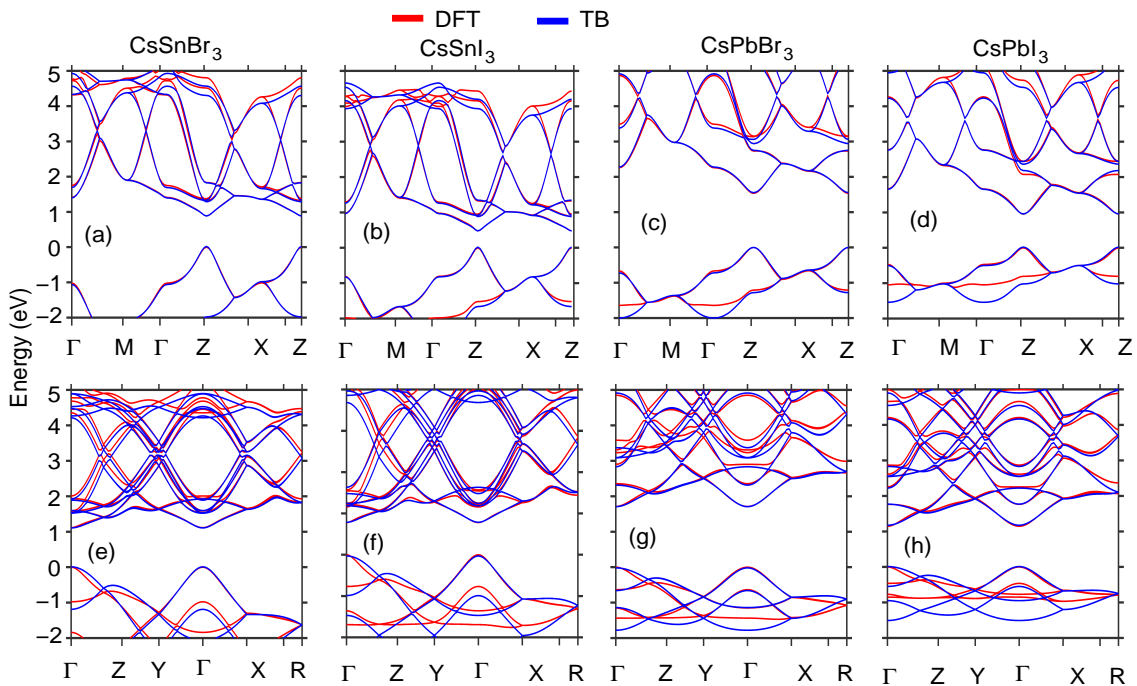}
\caption{ DFT fitted TB band structure of tetragonal and orthorhombic crystal phases of various halide perovskites. Here, Fermi level is set to zero.   }
\label{fig7}
\end{figure*}
\begin{figure}
\centering
\includegraphics[angle=-0.0,origin=c,height=8.5cm,width=9.0cm]{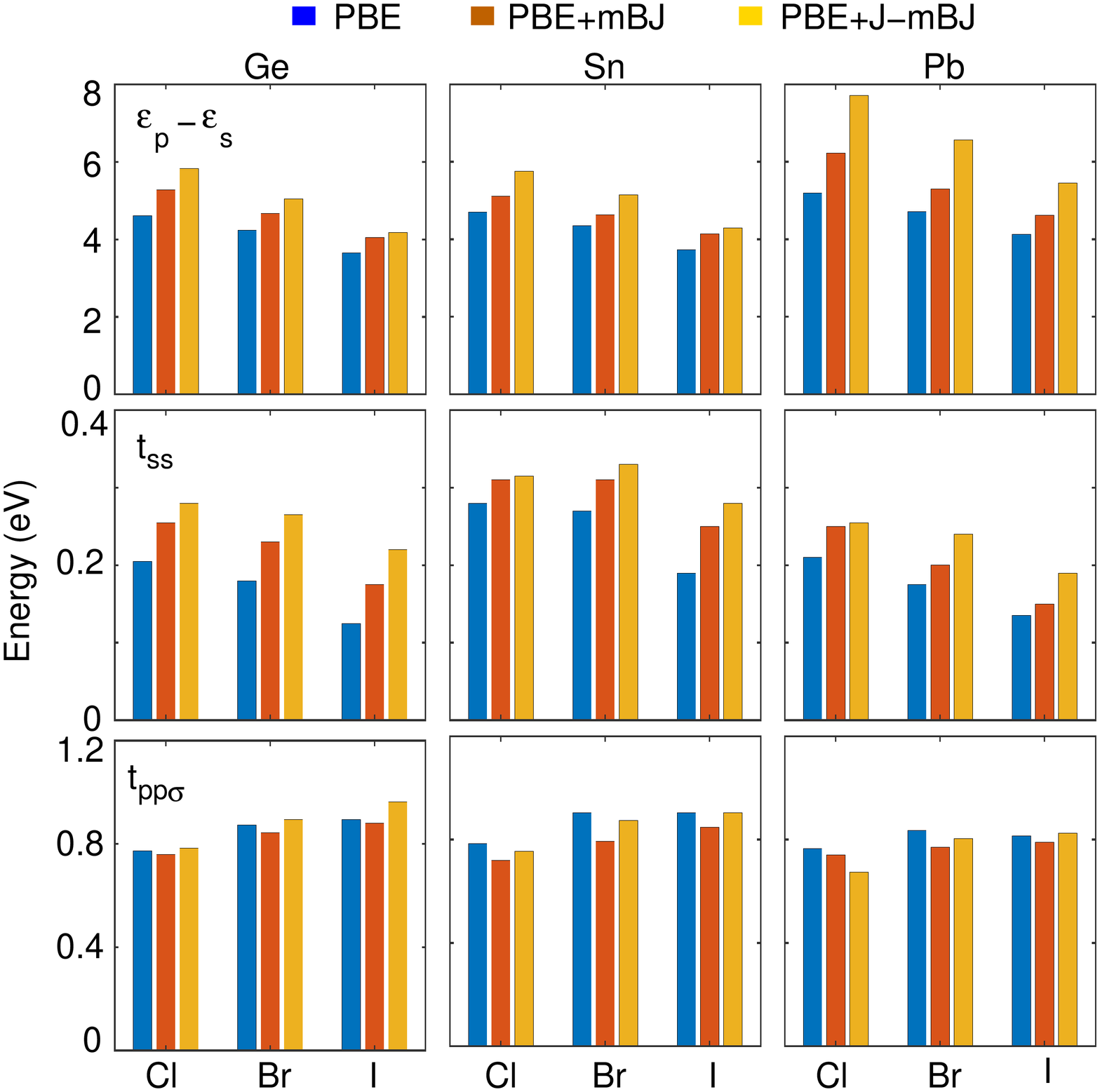}
\caption{ Comparison of various TB parameters of cubic halide perovskites obtained from fitting with the DFT band structure under various exchange-correlation functionals.}
\label{fig8}
\end{figure}

In the present section, we extend our minimal basis set TB model to tetragonal and orthorhombic systems. In the tetragonal phase, the in-plane rotation of the octahedra produces two in-equivalent B atoms.  We denote them as B$_A$ and B$_B$. Now the basis set include eight eigenstates, \textit{viz},  
$|s^{B_A}\rangle$, $|p^{B_A}_{x}\rangle$, $|p^{B_A}_{y}\rangle$, $|p^{B_A}_{z}\rangle$,$|s^{B_B}\rangle$, $|p^{B_B}_{x}\rangle$, $|p^{B_B}_{y}\rangle$, $|p^{B_B}_{z}\rangle$. The corresponding SOC included 16$\times$16. Thus, the spin-orbit coupled TB Hamiltonian is envisaged on 16 orbital basis set. For the complete Hamiltonian ref, \cite{Ravi}.  As the tetragonal phase having a pseudo-cubic structure and creates anisotropic interactions in in-plane and out of plane directions. Thus, there are two sets of TB parameters exist corresponding to in-plane and out of plane directions. The DFT fitted TB band structures of tetragonal phase for various halide perovskites are listed in Table \ref{T6}. As can be seen, the in-plane interactions' strength is weak compared to out of plane interactions. Similarly, in the case of the orthorhombic system, the rotation along $c$ direction creates inequivalence of B atoms, and thus the Hamiltonian matrix is designed for 32 orbitals basis set.  In Fig. \ref{fig7}, we have presented the DFT fitted TB band structure and fitting parameters are listed in Table \ref{T6}.The table infers that the interaction strength has decreased from that of the cubic phase, due to octahedral rotations.

\begin{table}[h]
\centering
\caption{ Interaction parameters ($\epsilon$ and $t$) for unstained equilibrium configurations in the units of eV. The SOC strength ($\lambda$) is estimated to be 0.14 eV. }
\begin{tabular}{cccccccc}
\hline
\hline
Phase  &  Interaction& $\epsilon_s$ & $\epsilon_p$ & $t_{ss}$ & $t_{sp}$& $t_{pp\sigma}$ & $t_{pp\pi}$ \\
  &   Path&  &  &  & &  &  \\
 \hline
$\beta-$CsSnBr$_3$  &$\hat{x}$,$\hat{y}$  & -1.52 & 2.99 &-0.27 &	0.47 &	0.8 &	0.09  \\
          &$\hat{z}$                      &  &3.08  &-0.25 &	0.49 &	0.72 &	0.09  \\\\
$\beta-$CsSnI$_3$  & $\hat{x}$,$\hat{y}$  & -1.24  & 2.59 &-0.21 &	0.45 &	0.8 &	0.085  \\
          &$\hat{z}$                      &  &2.64  &-0.21 &	0.43 &	0.74 &	0.08  \\\\
          
$\beta-$CsPbBr$_3$  &$\hat{x}$,$\hat{y}$  & -1  & 4.2 &-0.18 &	0.38 &	0.76 &	0.08  \\
          &$\hat{z}$                      &  & 4.24  &-0.16 &	0.38 &	0.66 &	0.07 \\\\
          
$\beta-$CsPbI$_3$  & $\hat{x}$,$\hat{y}$           & 0.76   & 3.54 &-0.13  &	0.38 &	0.74 &	0.08 \\
          &$\hat{z}$                      &    & 3.6 &-0.13 &	0.32  &	0.64 &	0.07 \\\\

$\gamma-$CsSnBr$_3$  &$\hat{x}$,$\hat{y}$             & -1.63 &	3.07 & -0.26&	0.4&	0.68&	0.07 \\
          &$\hat{z}$                      &  &  &-0.24	&0	&0.665&	0.06 \\\\

$\gamma-$CsSnI$_3$  & $\hat{x}$,$\hat{y}$           &-1.23	&2.66 &-0.18&	0.38&	0.64&	0.08\\
 &$\hat{z}$ &    &  &-0.24&	0&	0.665&	0.06 \\\\

$\gamma-$CsPbBr$_3$  &$\hat{x}$,$\hat{y}$             &-0.9&	4.23&-0.14&	0.4&	0.68&	0.05\\
&$\hat{z}$  & &    &-0.16&	0.04&	0.7&	0.07\\\\
          
$\gamma-$CsPbI$_3$  & $\hat{x}$,$\hat{y}$ & -0.75&3.59&-0.12&	-0.32&	0.64&	0.04\\
&$\hat{z}$  & &    &-0.14&	0.02&	0.66&	0.07\\          
          
\hline
\hline
\end{tabular}
\label{T6}
\end{table}

\subsection{Effect of Exchange-Correlation functionals on TB parameters.}

Having validated about the minimal basis set TB model, now we examine how the TB parameters are sensitive towards the XC functionals employed for DFT calculations. From the Eq. 9, the bandgap can be defined as E$_3$-E$_4$, which is thus the function of TB parameters. In the minimal basis set TB model, it turns out to be a E$_g$ = $E_p^{B} - E_s^{B} -2t_{pp\sigma}^{B}-4t_{pp\pi}^{B} +6t_{ss}^{B}$. It can be seen from the Fig. \ref{fig3} and \ref{fig99} that the bandgap of halide perovskites is highly sensitive to the type of exchange-correlation functionals used in performing DFT calculations. Thus, we would like to examine the the effect of XC functionals on these effective parameters. In Fig. \ref{fig8}, we have shown the effective TB parameters for three different XC \textit{viz.,} GGA, mBJ and Jishi-mBJ, functional for cubic halide perovskites to know their dependency on XC functionals. The figure infers that difference between the onsite energies increases with the bandgap and exhibit maximum values for Jishi-mBJ XC functional. Among all the hopping interactions, t$_{ss}$ influence is profound on the band spectrum and small change in the values affect the band topology significantly. 

\subsection{Role of $A$ Site Atom on the Electronic Structure.}
\begin{figure}
\centering
\includegraphics[angle=-0.0,origin=c,height=5.5cm,width=8.5cm]{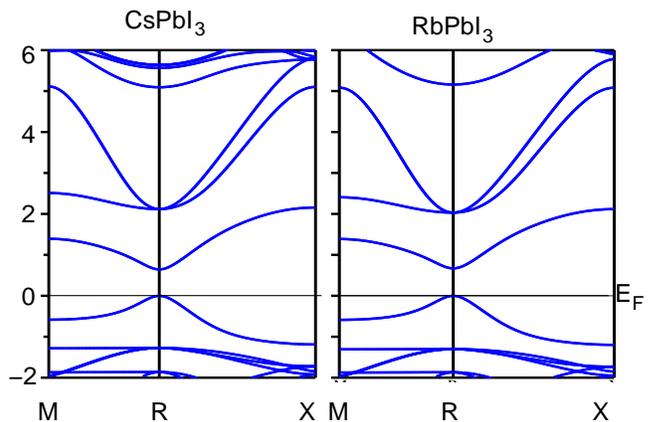}
\caption{ Band structure of CsPbI$_3$ and RbPbI$_3$ perovskites for the similar lattice parameter.}
\label{fig9}
\end{figure}

To know the role of $A$ site atoms, we have investigated the band structure of cubic RbPbX$_3$ by placing the Rb atom at Cs site without relaxing the structure and same is  shown in the Fig. \ref{fig9}. Our results show a tiny increase in the bandgap value of 0.02 eV for RbPbI$_3$ as compared with the CsPbI$_3$ and no significant changes have been observed in band spaghetti.  Structural relaxation carried out on the RbPbX$_3$ shows the reduced lattice parameters of the cubic lattice as compared to CsPbX$_3$. In contrast, the structural relaxation study with the organic molecule at A site shows the larger volume as compared to the CsPbX$_3$. However, the organic molecule introduces structural distortion in the octahedral cage, and compounds show noncentrosymmetric nature and this, in turn, affects the band structure profoundly. 

\subsection{Surface Band Structure}

\begin{figure*}
\centering
\includegraphics[angle=-0.0,origin=c,height=10cm,width=14.0cm]{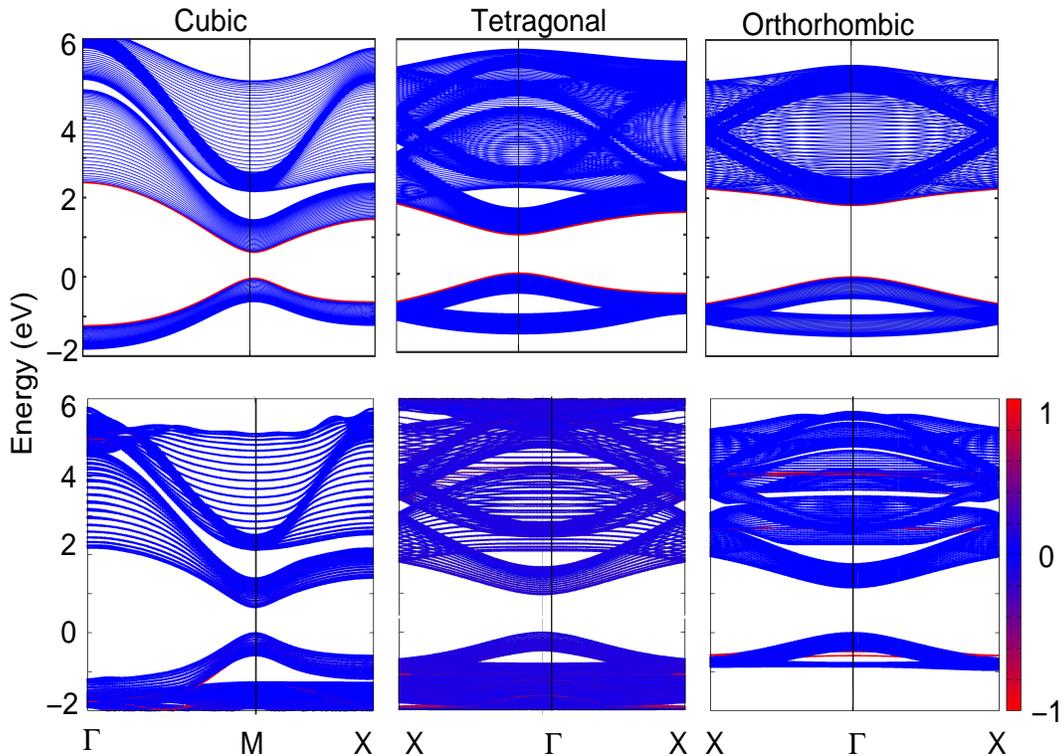}
\caption{ Surface band structure of CsPbI$_3$ polymorphs obtained from the SK-TB formalism (upper panel) and compared with band structure of Wannier function based TB model (lower panel). }
\label{fig10}
\end{figure*}

The bulk electronic structure study has enabled us to analyze the bulk properties of the halide perovskites. However, in many cases, the study requires the surface electronic structure to understand the surface and interface phenomena of these systems. The DFT based calculations on slab structure require a huge amount of computational time and memory to analyze these properties. Thus to overcome such difficulty, we build Slab TB model and obtain the surface electronic structure. The slab TB model is envisaged by taking four orbital basis of the proposed bulk TB model and according the surface bands are estimated from the bulk Hamiltonian by using the bulk TB parameters. 


Here, the slab consists of alternative stacking of CsI and BI$_2$ layers. However, as discussed in section-II, there are no contribution X-p orbitals at the Fermi level, the model is restricted to B-\{s, p\} orbital basis. The appropriate Hamiltonian for the slab consisting of $n$ unit cells along (001) direction. We employ bulk TB  parameters and construct the Hamiltonian in the desired direction. The matrix  form the TB Hamiltonian is given by 

\begin{equation}
  H_{TB}^{Slab} =\left (\begin{array}{cc}  
H_{\uparrow \uparrow}& H_{\uparrow \downarrow} \\
H_{\downarrow \uparrow}&  H_{\downarrow \downarrow}
  \end{array} \right).
\end{equation}
Where, 
\begin{equation}
    H_{\uparrow \uparrow} = 
    \left ( \begin{array}{cccccc} 
    H_{11} & H_{12} & 0 & 0 & 0&\dots \\
    H_{21} & H_{22} & H_{23} & 0 &0& \dots \\
    0 & H_{32} & H_{33} & H_{34} & 0&\dots \\
    \vdots & &\ddots&\ddots & \ddots \\
    \dots  & 0 &0&H_{n-1n-2}& H_{n-1n-1} & H_{n-1n}\\
    \dots & 0 & 0 &0& H_{nn-1} & H_{nn}
    \end{array} \right)
\end{equation}
Here, $H_{jj}$ is the Hamiltonian for $j^{th}$ layer of the slab,  and it is given by,
\begin{equation}
 H_{jj}= 
  \left( \begin{array}{cccc}  
    \epsilon_s+f_0 & 2it_{sp}^xS_x & 2it_{sp}^xS_y & 0\\
    -2it_{sp}^xS_x & \epsilon_{p}^x+f_1 & -i\lambda &0\\ 
    -2it_{sp}^xS_y & i\lambda & \epsilon_{p}^x+f_2 & 0\\
    0 &   0 & 0 & \epsilon_{p}^z+f_3 
\end{array}  \right)
\end{equation}
Here,
\begin{eqnarray}\label{3}
f_0& = &2t_{ss}^xcos(k_xa)+2t_{ss}^xcos(k_ya), \nonumber \\
f_1& = &2t_{pp\sigma }^xcos(k_xa)+2t_{pp\pi }^xcos(k_ya), \nonumber \\
f_2& = &2t_{pp\sigma }^xcos(k_ya)+2t_{pp\pi }^xcos(k_xa), \nonumber\\
f_3& = &2t_{pp\pi}^xcos(k_xa)+2t_{pp\pi }^xcos(k_ya).
\end{eqnarray}

The block H$_{jj-1}$ describe the interaction between layer j and j-1.  
\begin{equation}
H_{j-1j} = (H_{jj-1})^{T} = 
\left( \begin{array}{cccc}
        t_{ss}^z&0&0&t_{sp}^z\\
       0&t_{pp\pi}^z&0&0\\
    0&0&t_{pp\pi}^z&0\\  
    -t_{sp}^z&0&0&t_{pp\sigma}^z 
\end{array}  \right)
\end{equation}
The off diagonal block emerging from the SOC is of the form
\begin{equation}
    H_{\uparrow \downarrow}= 
    \left (\begin{array}{cccccc} 
    G_{11} & 0 & 0 & 0 & 0&\dots \\
    0 & G_{22} & 0 & 0 &0& \dots \\
    0 & 0& G_{33} & 0 & 0&\dots \\
    \vdots & &\ddots&\ddots & \ddots \\
    \dots  & 0 &0& 0&G_{n-1n-1} &0\\
    \dots & 0 & 0 &0& 0 & G_{nn}
    \end{array} \right )
    \end{equation}
    \begin{equation}
G_{jj}=\left( \begin{array}{cccc}
        0&0&0&0\\
    0&0&0&\lambda\\ 
    0&0&0&-i\lambda\\
    0&\lambda&-i\lambda&0
\end{array}  \right). 
\end{equation}

The band spectrum obtained from the diagonalization surface TB matrix for cubic, tetragonal and orthorhombic phases are shown in Fig. \ref{fig10} and is compared with Wannier functions based TB band structure.

\section{Conclusions}

In conclusion, the present work, we have analyzed the family of halide perovskites using both parameter free Density functional calculations and parametric tight-binding (TB) model through the Slater-Koster description. The analysis has unravelled the dominant orbital overlapping interactions that govern the bandgap and play an important role in exploring the optoelectronic applications and topological phases. Various exchange-correlation (XC) functionals (PBE, PBE+mBJ, HSE06, HSE and PBE+J-mBJ) were employed on nine halide perovskites in three structural polymorphs. While HSE and PBE+mBJ underestimate the bandgap by a similar magnitude, PBE+J-mBJ either gives bandgap close to the experimental value or overestimate it \cite{jishi1,Jishi2}. The corrections mBJ and J-mBJ adopts the same formalism, but with varying parameters. Furthermore, the study reveals that, though thirteen orbitals are involved in the chemical bonding, a four orbital based tight-binding model is good enough to capture energy dispersion in the momentum space in the vicinity of the Fermi level. The successful extension of the present minimal basis TB model to other lower symmetry crystal polymorphs and slab structure for various experimentally synthesized compounds provides the effectiveness of this model. Interestingly, the strength of the dominant electron hopping integrals are found to be nearly independent of the XC functional adopted and therefore, the proposed TB model has become more universal. Also, since it excellently reproduces the surface band structures, it can be further improved to study the transport phenomenon as the Wannier formalism does.

{\bf Acknowledgement:} The work is funded by  the Department of Science and Technology, India, through Grant No. CRG/2020/004330. We acknowledge the use of the computing resources at HPCE, IIT Madras.

{\bf Data availability:} The data that support the findings of this study are available from the corresponding author upon reasonable request.



\bibliography{paper}

\end{document}